\documentclass[aps,pre,twocolumn,groupedaddress]{revtex4-1}
\usepackage{amsmath,amssymb,graphicx,dsfont,tikz,subfigure,mathtools,color,float}
\usepackage[english]{babel}
\usepackage[utf8]{inputenc}
\usepackage{hyperref}

\newcommand{\bx}{\mathbf{x}}
\newcommand{\by}{\mathbf{y}}
\newcommand{\bbx}{\mathbf{\bar{x}}}
\newcommand{\bby}{\mathbf{\bar{y}}}
\newcommand{\bu}{\mathbf{u}}
\newcommand{\bv}{\mathbf{v}}
\newcommand{\bbu}{\mathbf{\bar{u}}}
\newcommand{\bbv}{\mathbf{\bar{v}}}
\newcommand{\bq}{\bar{q}}
\begin{document}
\title{Stochastic thermodynamics of interacting degrees of freedom: Fluctuation theorems for detached path probabilities}

\author{Jannik Ehrich}
\email{jannik.ehrich@uni-oldenburg.de}
\author{Andreas Engel}
\affiliation{Universit\"at Oldenburg, Institut f\"ur Physik, 26111 Oldenburg, Germany}
\date{\today}

\begin{abstract}
Systems with interacting degrees of freedom play a prominent role in stochastic thermodynamics. Our aim is to use the concept of detached path probabilities and detached entropy production for bipartite Markov processes and elaborate on a series of special cases including measurement-feedback systems, sensors and hidden Markov models. For these special cases we show that fluctuation theorems involving the detached entropy production recover known results which have been obtained separately before. Additionally, we show that the fluctuation relation for the detached entropy production can be used in model selection for data stemming from a hidden Markov model. We discuss the relation to previous approaches including those which use information flow or learning rate to quantify the influence of one subsystem on the other. In conclusion, we present a complete framework with which to find fluctuation relations for coupled systems.
\end{abstract}

\pacs{}
\maketitle

\section{Introduction}\label{sec:intro}
% Background
Within the field of stochastic thermodynamics~\cite{Seifert2012} systems with interacting degrees of freedom play an important role since they offer case studies for the interplay between thermodynamics and information theory~\cite{Parrondo2015}.

% Relevance
For such systems there are diverse setups. For example, systems with degrees of freedom which are inaccessible to the experimenter show deviations in the fluctuation relation for the entropy production~\cite{Mehl2012}. Further, there is a wide variety of measurement-feedback setups in which a stochastic system is measured and subsequently controlled by a feedback controller. Additionally, sensing is a manifestation of such a coupled joint system. Here, a stochastic system is influenced by an external stochastic process. Usually the task of the sensor is to measure the external process.

% State-of-the-art
Systems with interacting degrees of freedom have conveniently been modeled using bipartite Markov processes. For such systems a splitting of the second law of thermodynamics has been achieved such that individual second laws applicable to each one of the subsystems retain the influence of the other through information theoretic terms like information flow~\cite{Horowitz2014} or learning rate~\cite{Hartich2014}. Recently, Crooks and Still~\cite{Crooks2016} obtained marginal fluctuation relations applicable to one of two subsystems within a bipartite setup. In these the influence of the other subsystem is encoded in a transfer-entropy~\cite{Schreiber2000} term.

Measurement-feedback systems have been studied by Sagawa and Ueda~\cite{Sagawa2010,Sagawa2012}, Horowitz and Vaikuntanathan~\cite{Horowitz2010} and Ponmurugan~\cite{Ponmurugan2010}. They established fluctuation theorems applicable to such systems. Various sensor setups have been studied. Within the context of stochastic thermodynamics, the main focus is on the relation between the sensor's information about the external environment and its energy dissipation~\cite{Mehta2012,Lan2012,Barato2014,Sartori2014,Bo2015}.

% Knowledge gap
However, it is instructive to study these setups from a common perspective and find fluctuation relations which, when evaluated for these special cases, recover the known results. This also offers a formalism to reliably derive fluctuation relations applicable to one of several interacting subsystems.

% Paper objectives
The aim of this paper is to (1) put to use the concept of detached path probabilities and detached entropy production for bipartite Markov chains. Both have been introduced as preliminary quantities for further evaluation in~\cite{Crooks2016}; (2) elaborate on a series of special cases that have been studied separately before, namely measurement-feedback systems, sensors and hidden Markov models and show that fluctuation theorems involving the detached entropy production can recover known results; and (3) show how to use our formalism to confirm model parameters for hidden Markov models.

\section{Detached path probabilities}
\begin{figure}[ht]
 \centering
\includegraphics[width = \linewidth]{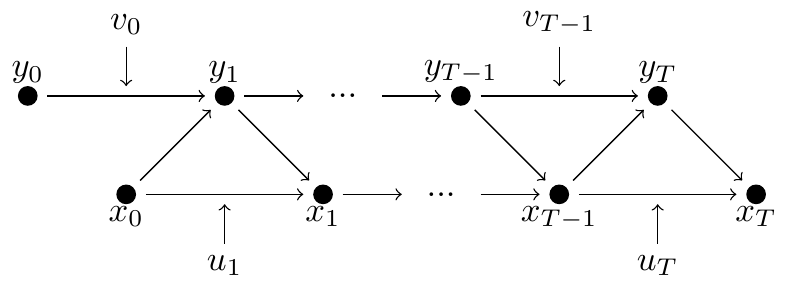}
 \caption{Setup of a joint trajectory. The subsystems $x$ and $y$ update one after the other.}
 \label{fig_setupJointProcess}
\end{figure}

We consider a two-variate Markov chain $(x_{0:T},y_{0:T})$ with 
\begin{subequations}
\begin{align}\label{defx}
 x_{0:T} &= \{x_0,x_1,...,x_T\}\eqqcolon\{x_0,\bx\}\\\label{defy}
 y_{0:T} &= \{y_0,y_1,...,y_T\}\eqqcolon\{y_0,\by\}.
\end{align}
\end{subequations}
The process is assumed to be bipartite as sketched in Fig.~\ref{fig_setupJointProcess} such that only one subsystem changes its state at a time \cite{Hartich2014,Horowitz2014,Crooks2016}. A simple example is given by two discretized coupled Langevin equations with independent noise sources. 

The stochastic dynamics depend on external protocols $\bu=\{u_1,...,u_T\}$ and $\bv=\{v_0,...,v_{T-1}\}$ that influence the individual transition probabilities $p_x(x_t|x_{t-1},y_t;u_t)$ and $p_y(y_t|x_{t-1},y_{t-1};v_{t-1})$, respectively. For the joint probability of the entire sequence of states, starting from an initial condition $(x_0,y_0)$, we hence have
\begin{multline}\label{defjointprob}
 p(\bx,\by|x_0,y_0;\bu,\bv) = p_y(y_1|x_0,y_0;v_0)\,p_x(x_1|x_0,y_1;u_1)...\\
            ...\times p_y(y_T|x_{T-1},y_{T-1};v_{T-1})\,p_x(x_T|x_{T-1},y_T;u_T).
\end{multline}
Together with the initial distribution $p_0(x_0,y_0)$ this probability uniquely determines the entire process.

From the structure of the transition probabilities as well as from Fig.~\ref{fig_setupJointProcess} it is clear that the stochastic variables $x$ and $y$ influence each other. Nevertheless, as observed by Crooks and Still~\cite{Crooks2016}, it is instructive to split the joint probability \eqref{defjointprob} according to 
\begin{align}
  p(\bx,\by&|x_0,y_0;\bu,\bv)= q_x(\bx|x_0; \by, \bu) \, q_y(\by|y_0;x_0,\bx,\bv),  \label{eqn_decompositionJointProbability}
\end{align}
where 
\begin{subequations}
\begin{align}\label{eqn_xTrajectoryProb}
  q_x(\bx|x_0;\by,\bu) &\coloneqq \prod\limits_{t=1}^{T} p_x(x_t|x_{t-1},y_t;u_t)\\
  q_y(\by|y_0;x_0,\bx,\bv) &\coloneqq \prod\limits_{t=1}^{T} p_y(y_t|x_{t-1},y_{t-1};v_{t-1}). \;  \label{eqn_yTrajectoryProb}
\end{align}
\end{subequations}
A related concept in information theory goes under the name of \emph{causal conditioning} and is applicable even to non-Markovian processes \cite{Marko1973, Massey1990, Jiao2013}.

\begin{figure}[ht]
 \centering
 \includegraphics[width=\linewidth]{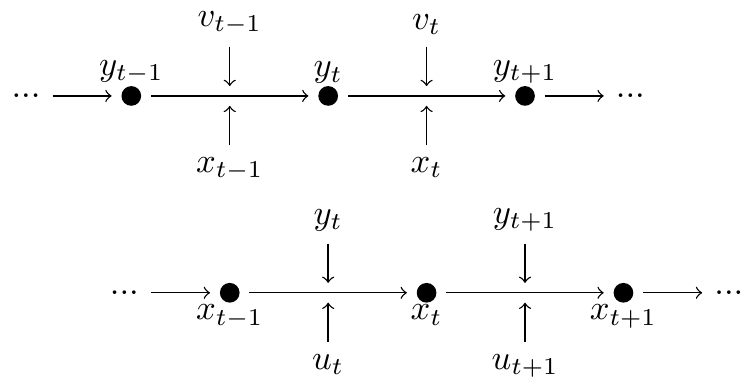}
 \caption{Decomposition of the joint trajectory into two sub-trajectories with the other process acting as a time-dependent protocol.}
 \label{fig_splitJointProcess}
\end{figure}

We call $q_x$ and $q_y$ the \emph{detached} path probabilities of the individual subsystems. They are normalized according to
\begin{align}
 1 &= \int d\bx\,q_x(\bx|x_0; \by, \bu)\label{eqn_detachedNormalized}\\
&= \int dx_1\,p_x(x_1|x_0,y_1;u_1) \, ... \int dx_T\,p_x(x_T|x_{T-1},y_T;u_t)\nonumber
\end{align}
and similarly for $q_y(\by|y_0;\bx,x_0,\bv)$. The decomposition of the joint probability into a product of detached probabilities according to Eq.~\eqref{eqn_decompositionJointProbability} can be represented graphically as a splitting of the joint process into two sub-processes in which the state of the other process enters as an additional time-dependent protocol. This is depicted in Fig.~\ref{fig_splitJointProcess}.

It is important to distinguish the detached path probabilities from the \emph{marginal} and the \emph{conditional} probabilities of the individual sequences. The marginal probability of the sequence of $x$-values is defined by 
\begin{align}\label{defmargprobx}
p_x(\bx|x_0;\bu,\bv)\coloneqq \int dy_0d\by \, p_0(y_0|x_0)\,p(\bx,\by|x_0,y_0;\bu,\bv),
\end{align}
where $p_0(y_0|x_0) = p_0(x_0,y_0)/\int dy_0\,p_0(x_0,y_0)$. Hence, in contrast to $q_x(\bx|x_0; \by, \bu)$, all dependence on $y$ is averaged out. Similarly, 
\begin{align}\label{defmargproby}
p_y(\by|y_0;\bu,\bv)\coloneqq \int dx_0d\bx \, p_0(x_0|y_0)\, p(\bx,\by|x_0,y_0;\bu,\bv)
\end{align}
with no trace left of the $x$-dynamics. 

The conditional probability of $\bx$ given the $y$-sequence, on the other hand, is defined by:
\begin{align}
 p_x(\bx|x_0, &y_0,\by; \bu,\bv)
= \frac{p(\bx,\by|x_0,y_0;\bu,\bv)}{\int d\bx\,p(\bx,\by|x_0,y_0;\bu,\bv)}\nonumber\\
 &=\frac{q_x(\bx|x_0;\by,\bu)\, q_y(\by|y_0;\bx,x_0,\bv)}{\int d\bx\, q_x(\bx|x_0;\by,\bu)\, q_y(\by|y_0;\bx,x_0,\bv)}.\label{diffqp}
\end{align}
Again analogous results hold for 
\begin{equation}
 p_y(\by|x_0, \bx,y_0; \bu,\bv) = \frac{p(\bx,\by|x_0,y_0;\bu,\bv)}{\int d\by\,p(\bx,\by|x_0,y_0;\bu,\bv)}.\label{defcondproby} 
\end{equation}
The conditional probability $p_x(\bx|x_0, y_0,\by; \bu,\bv)$ depends on the particular $y$-values considered. Yet, it is different from the detached probability $q_x(\bx|x_0; \by, \bu)$ since the latter ignores the feedback from $x$ to $y$. Formally, this implies that we cannot evaluate the integral in the denominator in Eq.~\eqref{diffqp}. More intuitively, the 
feedback between the subsystems implicates that a conditioning on future values of one trajectory constrains the evolution of the other one. It is precisely this effect that is eliminated in the definition of the detached path probabilities.

%%%%%%%%%%%%%%%%%%%%%%%%%%%%%%%%%%%%%%%%%%%%%%%%%%%%%%%%%%%%%%%%%%%%%%%%%%%%%%%%%%%%%%%%%%%%%%%%%%%%%%%%%%%%

\section{Entropy production and fluctuation relations}

\subsection{Reverse process}
Entropies together with the respective fluctuation theorems play a pivotal role in stochastic thermodynamics. Their definition generally involves a conjugate process \cite{Seifert2012}.

Although there is some freedom in choosing the conjugate process, here we will always take the time-reversed process driven by the time-reversed protocols $\bbu\coloneqq\{ u_T, u_{T-1},..., u_1\}$ and $\bbv\coloneqq\{ v_{T-1}, v_{T-2},..., v_0\}$ as the conjugate one. For simplicity and to lighten the notation, we assume that the protocol values $u$ and $v$ as well as the state variables $x$ and $y$ are even under time-reversal, i.e. $\bar u_t = u_t$, $\bar v_t = v_t$, $\bar x_t = x_t$ and $\bar y_t = y_t$ for all $t=0,...,T$. 

\begin{figure}[ht]
 \centering
 \includegraphics[width = \linewidth]{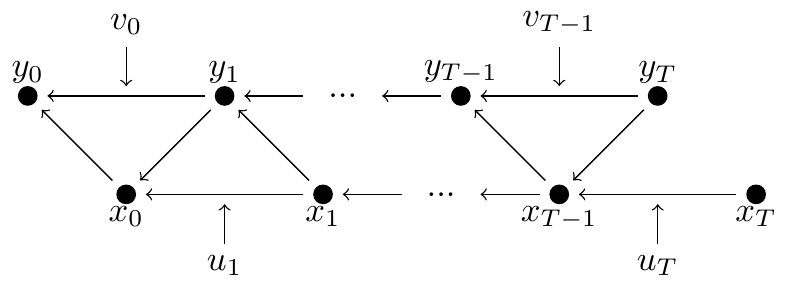}
 \caption{Reverse joint trajectory.}
 \label{fig_reverseJointProcess}
\end{figure}

The stochastic dynamics of the reversed process are characterized by a joint probability of the form \eqref{defjointprob} with $\bu$ und $\bv$ replaced by $\bbu$ and $\bbv$, respectively. We will need the path probabilities of the reversed process in particular for the time-reversed sequences of $x$ and $y$ states that we denote by
\begin{subequations} 
\begin{align}\label{defbarx}
 x_{T:0} &= \{x_T,x_{T-1},...,x_0\}\eqqcolon\{x_T,\bbx\}\\\label{defbary}
 y_{T:0} &= \{y_T,y_{T-1},...,y_0\}\eqqcolon\{y_T,\bby\}.
\end{align}
\end{subequations}
The joint probability of the reversed trajectory starting at $(x_T,y_T)$ under the reversed process is therefore given by
\begin{align}
 \bar p(\bbx,&\bby|x_T,y_T;\bbu,\bbv)\nonumber\\
 &= p_x(x_{T-1}|x_T,y_T,u_T)\,p_y(y_{T-1}|x_{T-1},y_T;v_{T-1}) ...\nonumber\\
 &\qquad\qquad \times p_x(x_0|x_1,y_1; u_1)\,p_y(y_0|x_0,y_1; v_0).\label{defjointprobrev}
\end{align}
This is shown graphically in Fig.~\ref{fig_reverseJointProcess}. Note that, compared to Fig.~\ref{fig_setupJointProcess}, horizontal components of arrows (indicating time) are reversed whereas vertical ones (indicating causal influence) remain unchanged \cite{Crooks2016}.

As initial distribution of the reverse process we take the final distribution of the forward process: 
\begin{align}
 p_T(x_T,y_T) \coloneqq p(x_T,y_T|x_0,y_0;\bu,\bv).
\end{align}

Analogous to Eq.~\eqref{eqn_decompositionJointProbability}, we decompose the joint probability of the reversed process into the respective detached probabilities~\cite{FootnoteReverseDecomposition}:
\begin{align}
  \bar p(\bbx,\bby|x_T,y_T;\bbu,\bbv)= \bq_x(\bbx|x_T; \bby, y_T,\bbu) \, \bq_y(\bby|y_T;\bbx,\bbv),\label{eqn_decompositionReverseJointProbability}
\end{align}
where now 
\begin{subequations}
\begin{align}\label{eqn_xReversedTrajectoryProb}
  \bq_x(\bbx&|x_T;\bby,y_T,\bbu) \coloneqq \prod\limits_{t=1}^{T} p_x(x_{t-1}|x_t,y_t; u_t)\\
  \bq_y(\bby&|y_T;\bbx,\bbv) \coloneqq \prod\limits_{t=1}^{T} p_y(y_{t-1}|x_{t-1},y_t; v_{t-1}) \; . \label{eqn_yReversedTrajectoryProb} 
\end{align}
\end{subequations}

\subsection{Joint entropy production}

With the specification of the reverse process, we may define the corresponding entropy productions. It is natural to consider the \emph{joint} entropy production 
\begin{align}
 \sigma_{xy} \coloneqq 
     \ln \frac{p_0(x_0,y_0)\,p(\bx,\by|x_0,y_0;\bu,\bv)}
                {p_T(x_T,y_T)\,\bar p(\bbx,\bby|x_T,y_T;\bbu,\bbv)}\label{eqn_defJointEntropy}\; .
\end{align} 
As usual $\sigma_{xy}$ quantifies the relative surprise to observe the time-reversed trajectory 
$\{x_T,\bbx,y_T,\bby\}$ under the influence of the time-reversed protocols $\{\bbu,\bbv\}$ if the forward trajectory $\left\{x_0,\bx,y_0,\by\right\}$ was seen under the protocols $\left\{\bu,\bv\right\}$. 

To avoid clutter, we will from now on suppress the dependence on the protocols $\bu$ and $\bv$ and their reversed versions and show it only where it matters.

The joint entropy production fulfills an integral fluctuation theorem (IFT):
\begin{align}
 \left\langle e^{-\sigma_{xy}}\right\rangle &= \int d\bx d\by dx_0 dy_0 \,e^{-\sigma_{xy}}\, p_0(x_0,y_0)\,p(\bx,\by|x_0,y_0)\nonumber\\
  &= \int d\bbx d\bby dx_T dy_T \, p_T(x_T,y_T)\,\bar p(\bbx,\bby|x_T,y_T)\nonumber\\
 &=1.\label{eqn_jointIFT}
\end{align}

\subsection{Detached entropy production}

Building on the detached path probabilities we may define the \emph{detached} entropy productions
\begin{subequations}
\begin{align}\label{appentx}
 \tilde\sigma_x &\coloneqq \ln\frac{p_0(x_0|y_0)\,q_x(\bx|x_0;\by)}{p_T(x_T|y_T)\,\bq_x(\bbx|x_T;y_T,\bby)}
            \\\mathrm{and}\nonumber &\\\label{appenty}
 \tilde\sigma_y &\coloneqq \ln\frac{p_0(y_0|x_0)\,q_y(\by|y_0;x_0,\bx)}{p_T(y_T|x_T)\,\bq_y(\bby|y_T;\bbx)}\; , 
\end{align}
\end{subequations}
where the initial conditions may be calculated from $p_0(x_0,y_0)$ and $p_T(x_T,y_T)$, respectively.

Using~Eq.~\eqref{appentx}, we obtain:
\begin{align}
 \left\langle e^{-\tilde\sigma_{x}}\right\rangle &= \int d\bx d\by dx_0 dy_0 \,\frac{p_T(x_T|y_T)\,\bar q_x(\bbx|x_T;y_T,\bby)}{p_0(x_0|y_0)\,q_x(\bx|x_0;\by)}\nonumber\\
 &\qquad\qquad \times p_0(x_0,y_0)\,p(\bx,\by|x_0,y_0)\\
 &= \gamma,\label{eqn_detachedIFTx}
\end{align}
where, with Eq.~\eqref{eqn_decompositionJointProbability},
\begin{align}
 \gamma &\coloneqq \int d\bx d\by dx_0 dy_0 \, \left[ p_T(x_T|y_T)\,\bar q_x(\bbx|x_T;y_T,\bby)\right] \nonumber\\
&\qquad\qquad \times \left[ p_0(y_0)\,q_y(\by|y_0;x_0,\bx) \right] \label{eqn_defGamma}
\end{align}
is a parameter related to feedback from $x$ to $y$. In general $\gamma \neq 1$. Only if there is no feedback from $x$ to $y$, $q_y(\by|y_0;x_0,\bx) =  p_y(\by|y_0)$, we find
\begin{align}
 \gamma &= \int d\by dy_0 \, p_0(y_0)\,p_y(\by|y_0) \nonumber\\
 &\qquad \times \int d\bbx dx_T\, p_T(x_T|y_T)\,\bar q_x(\bbx|x_T;y_T,\bby) = 1 \label{eqn_gammaEqual1}.
\end{align}
In general, therefore, the detached entropy productions do not fulfill IFTs of the usual type.

\subsection{Relations between entropy productions}

The entropy productions defined above are not independent of each other. With the help of the mutual information \cite{Cover2006}
\begin{align}
 i(x,y)&\coloneqq \ln\frac{p(x|y)}{p(x)} = \ln\frac{p(y|x)}{p(y)}\label{eqn_defMutualInfo}
\end{align} 
we may disentangle the initial distribution according to 
\begin{equation}
 p_0(x_0,y_0)=p_0(x_0|y_0)\,p_0(y_0|x_0)\,e^{-i_0(x_0,y_0)}
\end{equation} 
and analogously for $p_T(x_T,y_T)$. We then find from Eqs. \eqref{eqn_decompositionJointProbability}, \eqref{eqn_decompositionReverseJointProbability}, \eqref{eqn_defJointEntropy}, \eqref{appentx}, and \eqref{appenty}
\begin{equation}\label{eqn_splitJointEntropy}
 \sigma_{xy}=\tilde\sigma_x+\tilde\sigma_y+\Delta i
\end{equation} 
where 
\begin{equation}\label{eqn_defDeltaI}
 \Delta i =i_T(x_T,y_T)-i_0(x_0,y_0)\; .
\end{equation} 

The IFT~\eqref{eqn_jointIFT} for the joint entropy production as well as the IFT~\eqref{eqn_detachedIFTx} for the detached entropy production together with the relation~\eqref{eqn_splitJointEntropy} between the entropy productions are the main general results of this study. We now specify the setup depicted in Fig.~\ref{fig_setupJointProcess} to various particular situations and elucidate the relation between the IFTs \eqref{eqn_jointIFT} and \eqref{eqn_detachedIFTx} and results obtained in previous studies of these systems.

\section{Special Cases}

\subsection{Measurement-feedback systems}
A measurement-feedback system consists of a system in contact with a thermal reservoir and a controller which measures the state of the system and uses that information to influence it. Most prominent examples of this type are Maxwell's demons and information engines~\cite{Parrondo2015}.

\begin{figure}[ht]
 \centering
 \includegraphics[width = \linewidth]{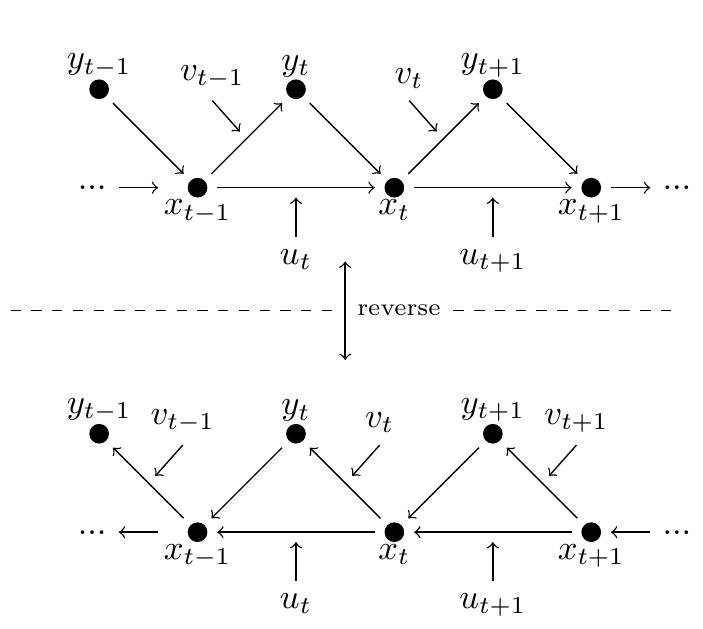}
 \caption{Setup of a measurement-feedback system. Top: the system $x$ is under feedback control from the controller $y$. In every timestep, $y$ represents a measurement of the current state of $x$ which is then used to influence its dynamics. Bottom: time-reversed process.}
 \label{fig_measFBSetup}
\end{figure}

We adapt our framework to this situation by describing the system by $x$ and the controller by $y$ (see Fig.~\ref{fig_measFBSetup}). Since subsystem $x$ is coupled to a thermal reservoir at constant inverse temperature $\beta$, the detached entropy production $\tilde\sigma_x$ of Eq.~\eqref{appenty} has, due to detailed balance, a clear thermodynamic interpretation~\cite{Seifert2005}:
\begin{align}
 \tilde\sigma_x &= \ln\frac{p_0(x_0|y_0)}{p_T(x_T|y_T)}\nonumber\\
&\;\;\;+\ln\frac{p_x(x_1|x_0,y_1;u_1)\,...\,p_x(x_T|x_{T-1},y_T;u_T)}{p_x(x_{T-1}|x_T,y_T;\bar u_T)\,...\,p_x(x_0|x_{1},y_1; \bar u_1)}\\
&= \ln\frac{p_0(x_0|y_0)}{p_T(x_T|y_T)}\nonumber\\
&\;\;\;-\beta\left[ H_x(x_1;y_1,u_1) - H_x(x_0;y_1,u_1) + ... \right.\nonumber\\
&\qquad\;\;\;\; \left. + H_x(x_T;y_T,u_T)-H_x(x_{T-1};y_T,u_T) \right]\\\label{eq:h1}
&\eqqcolon \Delta s_{x|y} -\beta\, Q_x.
\end{align}
Here $H_x(x;y,u)$ is the Hamiltonian governing the $x$-dynamics, $\Delta s_{x|y}$ is the change in conditional system entropy, and $Q_x$ denotes the heat that the system exchanges with the reservoir.

The evolution of the $y$-subsystem is solely determined by the sequential measurements it takes of the state of the $x$-subsystem:
\begin{align}
 p_y(y_{t+1}|x_t,y_t;v_t) = p_y(y_{t+1}|x_t;v_t).
\end{align}
Graphically, this is expressed by the absence of horizontal arrows in the $y$-trajectory in Fig.~\ref{fig_measFBSetup}. Moreover, the noise in the controller is \emph{measurement} noise and has no direct thermodynamic interpretation. The role of the protocol $v_t$ can be understood as influence on the measurement procedure, e.g. controlling its variance.

The detached entropy production $\tilde\sigma_y$ of Eq.~\eqref{appenty} then simplifies to:
\begin{align}
 \tilde\sigma_y &= \ln\frac{p_0(y_0|x_0)\,\prod\limits_{t=1}^T p_y(y_t|x_{t-1};v_{t-1})}{p_T(y_T|x_T)\,\prod\limits_{t=1}^T p_y(y_{t-1}|x_{t-1};v_{t-1})}\\
&=\ln\frac{p_0(y_0|x_0)}{p_T(y_T|x_T)} + \sum\limits_{t=1}^T \ln\frac{p(y_t|x_{t-1})}{p(y_{t-1}|x_{t-1})}\\
&= \sum\limits_{t=1}^T \ln \frac{p(y_t|x_{t-1})}{p(y_t|x_t)}
\end{align}
which upon using Eq.~\eqref{eqn_defMutualInfo} becomes:
\begin{align}
 \tilde\sigma_y &= -\sum\limits_{t=1}^T \ln\frac{p(y_t|x_t)}{p(y_t)} - \ln\frac{p(y_t|x_{t-1})}{p(y_t)}\\
&= -\sum\limits_{t=1}^T i(x_{t},y_t) - i(x_{t-1},y_{t})\\
&\eqqcolon -i_x.\label{eqn_detachedEntMeasFB}
\end{align}
Here $i_x$ denotes the \emph{information flow}~\cite{Horowitz2014}, which is the change in mutual information between $x$ and  $y$ that is caused by the changes in $x$ only. Conversely, 
\begin{equation}\label{eqn_infoFlowy}
 i_y:=\sum\limits_{t=1}^T i(x_{t-1},y_{t}) - i(x_{t-1},y_{t-1})
\end{equation} 
is the part of the mutual information that is solely due to the $y$-dynamics. Clearly,
\begin{align}
 i_x + i_y = \Delta i.\label{eqn_relationInfoFlows}
\end{align}
From Eqs.~\eqref{eqn_splitJointEntropy}, \eqref{eq:h1}, and \eqref{eqn_detachedEntMeasFB}  the joint entropy production~\eqref{eqn_defJointEntropy} reads:
\begin{align}
 \sigma_{xy} &= \Delta s_{x|y} -\beta Q_x-i_x+\Delta i\\
&= \Delta s_{x|y} -\beta Q_x + i_y\; .
\end{align}

The joint IFT~\eqref{eqn_jointIFT} acquires the form
\begin{align}
 \left\langle e^{-\Delta s_{x|y} + \beta Q_x - i_y} \right\rangle = 1, \label{eqn_jointIFTMeasFB}
\end{align}
which is a special version of the \emph{generalized Jarzynski equality}~\cite{Sagawa2010,Ponmurugan2010,Horowitz2010,Sagawa2012} for measurement-feedback processes: the entropy production of the feedback-controlled system needs to be augmented by an information quantity to fulfill the fluctuation theorem. 

Without this additional term the right-hand-side of the fluctuation relation for the entropy production $\tilde\sigma_x$ of the system alone deviates from unity, cf. Eq.~\eqref{eqn_detachedIFTx}:
\begin{align}
 \left\langle e^{-\Delta s_{x|y} - \beta Q_x} \right\rangle = \gamma. 
\end{align}
The parameter $\gamma$ defined in Eq.\eqref{eqn_defGamma} transforms to:
\begin{align}
  \gamma &\coloneqq \iint d\bx d\by dx_0 dy_0 \, \left[ p_T(x_T|y_T)\,\bar q_x(\bbx|x_T;y_T,\bby)\right] \nonumber\\
&\qquad\qquad \times \left[ p_0(y_0)\,\prod\limits_{t=1}^T p_y(y_{t}|x_{t-1}; v_{t-1}) \right],
\end{align}
and coincides with the ``efficacy parameter'' found by Sagawa and Ueda~\cite{Sagawa2012}. In our case, the time-reversed measurements are similar to the forward measurements because of the assumption that all variables are even under time-reversal.

\subsection{Sensors}
\begin{figure}[ht]
 \centering
 \includegraphics[width = 1\linewidth]{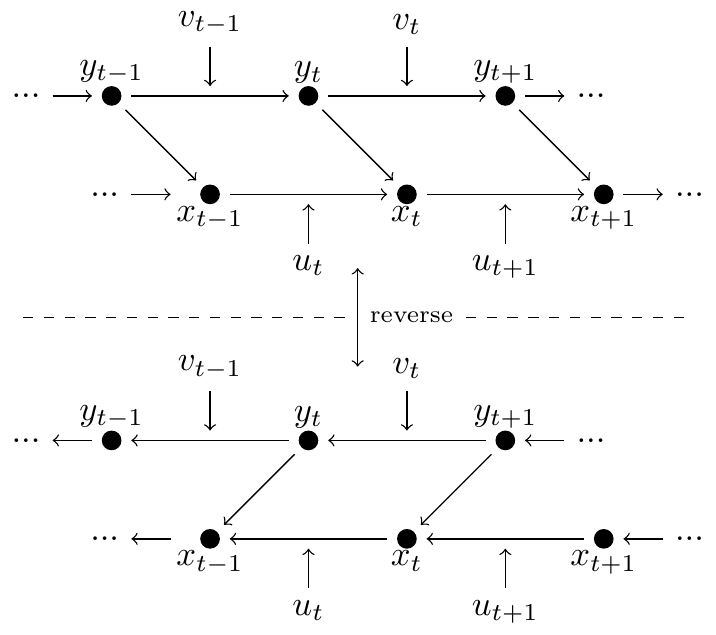}
 \caption{Setup of a sensor. Top: the sensor $x$ is influenced by the environment $y$ which is modeled as a stochastic process. Bottom: time-reversed process.}
 \label{fig_sensorSetup}
\end{figure}

Another setup involving the interplay between thermodynamics and information is sensing~\cite{Parrondo2015}. Accordingly, it has already been studied extensively within the framework of stochastic thermodynamics~\cite{Lan2012,Mehta2012,Still2012,Barato2013,Sartori2014,Barato2014,Ito2015,Bo2015,Hartich2016}. Here, a system is tasked with measuring an external signal. In the case of a biomolecular sensor, the external signal might be a chemical concentration, the pH, or osmotic pressure. The key ingredient is the fact that these signals are by themselves random and may be modeled by a stochastic process. We adapt our general setup to fit this situation by eliminating the feedback from $x$ to $y$, see Fig.~\ref{fig_sensorSetup}. Consequently, the $x$-subsystem acts as the sensor that measures the external process $y$.

Note that this is the biologically relevant interpretation of a \emph{sensor} since any molecular reaction system retains at least some memory of its past and is, in that sense, not an optimal sensor.

As in the previous example, the noise in the $x$-subsystem is of thermal nature and we have, cf. Eq.~\eqref{eq:h1}
\begin{align}\label{eqn_detachedEntSensor}
 \tilde\sigma_x = \Delta s_{x|y} - \beta Q_x.
\end{align}

Since $y$ remains unaffected by $x$, the detached path probability $q_y$~[Eq.~\eqref{eqn_yTrajectoryProb}] equals the marginal path probability $p_y$~[Eq.~\eqref{defcondproby}]:
\begin{align}
 q_y(\by|y_0;x_0,\bx) = p_y(\by|y_0)\label{eqn_sensorDetachedyEqualMarginaly}
\end{align}
and similarly for the reverse process:
\begin{align}
 \bar q_y(\bby|y_T;\bbx) = \bar p_y(\bby|y_T),
\end{align}
where $p_y(\bby|y_T)$ is the reverse marginal probability of the $y$-process.

This implies that, apart from the initial and final states, the detached entropy production $\tilde\sigma_y$ does not depend on the specific trajectory traversed by $x$ and is therefore related to the usual entropy production $\sigma_y$ of the $y$-process. Using Eqs.~\eqref{eqn_defMutualInfo} and~\eqref{eqn_defDeltaI} we find:
\begin{align}
 \tilde\sigma_y &= \ln\frac{p_0(y_0|x_0)\,p_y(\by|y_0)}{p_T(y_T|x_T)\,\bar p_y(\bby|y_T)}\\
&= \ln\frac{p_0(y_0)\,p_y(\by|y_0)}{p_T(y_T)\,\bar p_y(\bby|y_T)} -\Delta i\\
&\eqqcolon \sigma_y -\Delta i.\label{eqn_sensorDetachedEPyEqualMarginaly}
\end{align}

With Eqs.~\eqref{eqn_splitJointEntropy} and~\eqref{eqn_detachedEntSensor} the joint IFT now reads
\begin{align}
 \left\langle e^{-\Delta s_{x|y} + \beta Q_x -\sigma_y } \right\rangle = 1.
\end{align}
Jensen's inequality implies 
\begin{align}\label{eq:h3}
 \left\langle \Delta s_{x|y} \right\rangle - \beta \left\langle Q \right\rangle \geq - \left\langle \sigma_y \right\rangle,
\end{align}
which means that the average dissipation of the sensor is bounded from below by the negative entropy production of the external process. This relation has been shown in the steady state and for entropy rates in~\cite{Hartich2014}.

Since there is no feedback from $x$ to $y$ in this setup, the right-hand-side of the detached IFT given in Eq.~\eqref{eqn_detachedIFTx} equals 1, as we have shown in Eq.~\eqref{eqn_gammaEqual1}:
\begin{align}
\left\langle e^{-\Delta s_{x|y}+ \beta Q_x} \right\rangle = 1.\label{eqn_detachedIFTSensor}
\end{align}
For the sensor the usual fluctuation theorem thus holds if the conditional system entropy is used in the definition of the entropy production. This measure of dissipation has been applied to a nonequilibrium sensor in~\cite{Still2012}. If one wants to retain the usual definition of marginal system entropy change $\Delta s_x$, Eq.~\eqref{eqn_detachedIFTSensor} implicates a lower bound on the system's dissipation:
\begin{align}
 \left\langle \Delta s_x \right\rangle - \beta \left\langle Q \right\rangle \geq \left\langle \Delta i \right\rangle,
\end{align}
where we have used: $s_{x|y} = s_x - i(x,y)$. The average total sensor dissipation is therefore bounded from below by the change in mutual information it managed to build up during the process. In the context of sensory adaptation, this bound has been obtained and further analyzed by Sartori \emph{et al.}~\cite{Sartori2014}. We showed here that it can be deduced from a specialization of the detached fluctuation theorem.

\subsection{Hidden Markov models}
\begin{figure}[ht]
 \centering
 \includegraphics[width = \linewidth]{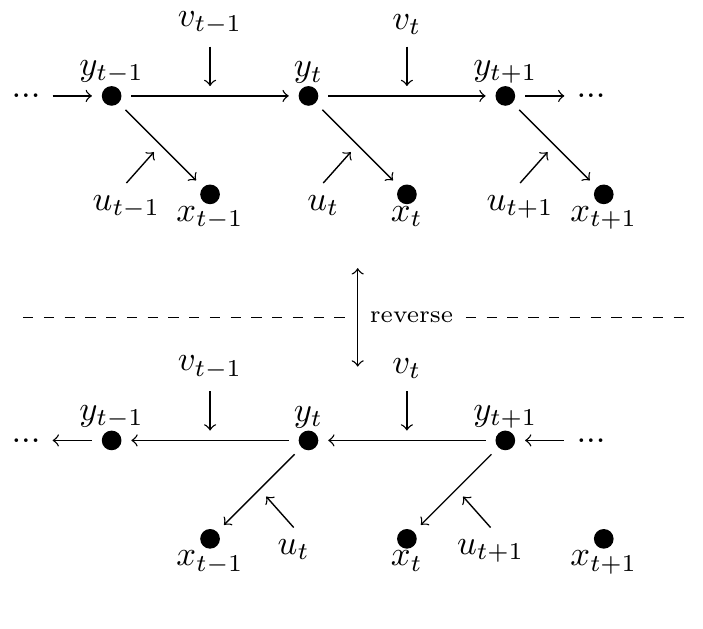}
 \caption{Setup of a hidden Markov model. Top: $y$ is a stochastic process that can only be observed through noisy measurements $x$. Bottom: time-reversed process.}
 \label{fig_hmmSetup}
\end{figure}

Hidden Markov models~\cite{Bishop2006} are a tool used, e.g., in machine learning to model sequential data coming from a Markov chain that is not directly accessible. Instead, one observes only measurements of its hidden states. Bechhoefer~\cite{Bechhoefer2015} has used the formalism of hidden Markov models to clarify feedback schemes in stochastic thermodynamics. Phrased in our setup the hidden Markov model appears as a special case of a sensor that has no memory of its past, cf. Fig.~\ref{fig_hmmSetup}.

Compared to the sensor, the $y$-dynamics remain unmodified. However, since the $x$-dynamics consist of measurements, the detached entropy production $\tilde\sigma_x$ simplifies in much the same way as the $y$-entropy production in the setup of a measurement-feedback system:
\begin{align}
 p_x(x_t|x_{t-1},y_t;u_t) = p_x(x_{t}|y_t; u_t).
\end{align}
We obtain, with Eqs.~\eqref{appentx}, \eqref{eqn_defMutualInfo}, and~\eqref{eqn_infoFlowy}
\begin{align}
 \tilde\sigma_x &= \ln\frac{p_0(x_0|y_0)}{p_T(x_T|y_T)}+ \sum\limits_{t=1}^T \ln\frac{p_x(x_t|y_t;u_t)}{p_x(x_{t-1}|y_t;u_t)}\\
&= \sum\limits_{t=1}^T \ln\frac{p(x_{t-1}|y_{t-1})}{p(x_{t-1}|y_t)}\\
&= \sum\limits_{t=1}^T i(x_{t-1},y_{t-1}) - i(x_{t-1},y_t)\\
&= -i_y,\label{eqn_HMMDetachedEPx}
\end{align}
where, as before, $i_y$ is the information flow due to the $y$-dynamics.

Due to the lack of feedback from $x$ to $y$, the apparent entropy production $\tilde\sigma_y$ equals the marginal entropy production minus the change in mutual information as in Eq.~\eqref{eqn_sensorDetachedEPyEqualMarginaly}.

The joint entropy production of Eq.~\eqref{eqn_splitJointEntropy} is therefore given by:
\begin{align}
 \sigma_{xy} = -i_y + \sigma_y.
\end{align}

Thus the joint IFT reads:
\begin{align}
\left\langle e^{-\sigma_y + i_y} \right\rangle = 1,
\end{align}
which is similar to the joint IFT for measurement-feedback processes~[Eq.\eqref{eqn_jointIFTMeasFB}]. In this case the roles of $x$ and $y$ are exchanged and the entropy production $\sigma_y$ needs not necessarily have a thermodynamic interpretation. 

The detached IFT~\eqref{eqn_detachedIFTx} yields:
\begin{align}\label{eqn_detachedIFTHMM}
\left\langle e^{i_y} \right\rangle  = 1.
\end{align}

This is an integral fluctuation theorem involving an informational quantity characterizing the correlation between hidden ($y$) and observed ($x$) process. It allows us to directly assess the direction of the information flow in the process. It implies $\langle i_y \rangle \leq 0$, meaning that on average information flows from the hidden process to the observations. This holds for arbitrary initial conditions. The larger the information flow, the more predictive the current observation is of the future hidden state.

In real-world applications, however, the computation of the information flow $i_y$ requires the knowledge of the observed sequence as well as the hidden sequence, which by definition is not accessible. To evaluate Eq.~\eqref{eqn_detachedIFTHMM} when only the observed sequence is available, one first needs to average over all possible hidden sequences:
\begin{align}
 \left\langle e^{i_y} \right\rangle = \left\langle \left\langle e^{i_y} \right\rangle_{p(y_0,\by|x_0,\bx)} \right\rangle_{p_x(x_0,\bx)} = 1. \label{eqn_HMMModelVerification}
\end{align}
In this way the detached IFT can be used for model verification given a sufficiently large data set of $x$-trajectories as we demonstrate in the following. For this we provide the auxiliary entropy production $\bar\sigma_x[x_0,\bx]$ simply by redefining the quantity in the average in Eq.~\eqref{eqn_HMMModelVerification}:
\begin{align}
 e^{-\bar\sigma_x[x_0,\bx]} &\coloneqq \left\langle e^{i_y[x_0,y_0,\bx,\by]} \right\rangle_{p(y_0,\by|x_0,\bx)}\\
&= \iint dy_0 d\by \; p(y_0,\by|x_0,\bx) \, \prod\limits_{t=1}^{T} \frac{p(x_{t-1}|y_{t})}{p(x_{t-1}|y_{t-1})},\label{eqn_defAuxEnt}
\end{align}
where $p(y_0,\by|x_0,\bx)$ is the \textit{posterior distribution} of the entire sequence $\{y_0,\by\}$ of hidden states given the entire sequence $\{x_0,\bx\}$ of observed states. The details of how this distribution can be calculated are given in the appendix.

Now we can assign to each observed trajectory $\{x_0,\bx\}$ an auxiliary entropy production $\bar\sigma_x[x_0,\bx]$. From Eq.~\eqref{eqn_HMMModelVerification} we infer that this entropy production fulfills an IFT:
\begin{align}\label{eqn_IFTAuxEnt}
 \left\langle e^{-\bar \sigma_x[x_0,\bx]} \right\rangle_{p_x(x_0,\bx)} = 1.
\end{align}

Thus, if we are given a set of trajectories and we want to infer whether the data have been generated from some specific model $\{p_y(y_{t+1}|y_t), p_x(x_t|y_t), p_0(y_0)\}$, we can calculate the auxiliary entropy production $\bar \sigma_x[x_0,\bx]$ for each trajectory and verify that the IFT in Eq.~\eqref{eqn_IFTAuxEnt} holds. If it does not, the model is not correct.

\subsubsection*{Example}

To illustrate this procedure, we consider a hidden Markov model with two hidden and two observed states each labeled with 0 and 1. The transition probabilities for the hidden Markov chain are given by the transition matrix $\tilde{\mathbb{T}}_y$:
\begin{align}
\tilde{\mathbb{T}}_y &\coloneqq \begin{pmatrix}
p_y(y_{t+1} = 0|y_t = 0) & p_y(y_{t+1} = 0|y_t = 1) \\
p_y(y_{t+1} = 1|y_t = 0) & p_y(y_{t+1} = 1|y_t = 1)
\end{pmatrix}\\ 
&= \begin{pmatrix}
1-a & b \\
a & 1-b
\end{pmatrix}.
\end{align}

The observed sequence is uniquely determined by the hidden sequence. For each time step the following holds:
\begin{align}
 p_x(x_t = y_t|y_t) = 1-\epsilon\qquad p_x(x_t = 1 - y_t|y_t) = \epsilon.
\end{align}
Thus each measurement is wrong with probability $\epsilon$. The transition probabilities for the observed chain is given by the transition matrix $\tilde T_x$:
\begin{align}
 \tilde{\mathbb{T}}_x &\coloneqq \begin{pmatrix}
p_x(x_{t} = 0|y_t = 0) & p_x(x_{t} = 0|y_t = 1) \\
p_x(y_{t} = 1|y_t = 0) & p_x(x_{t} = 1|y_t = 1)
\end{pmatrix} \\
&= \begin{pmatrix}
1-\epsilon & \epsilon \\
\epsilon & 1-\epsilon
\end{pmatrix}.
\end{align}

The full bipartite process is composed of the four states $(x,y) = (0,0), (0,1), (1,0)$, and $(1,1)$ in that order. It is described by the transition matrices $\mathbb{T}_y$ and $\mathbb{T}_x$ for the $y$- and $x$-step separately:
\begin{align}
\mathbb{T}_y &\coloneqq \mathbb{P}(x_t,y_{t+1}|x_t,y_t) = \begin{pmatrix}
1-a & b & 0 & 0 \\
a & 1-b & 0 &0 \\
0 & 0 & 1-a & b \\
0 & 0 & a & 1-b
\end{pmatrix}\\
\mathbb{T}_x &\coloneqq \mathbb{P}(x_{t+1},y_{t+1}|x_t,y_{t+1})\nonumber\\ 
&= \begin{pmatrix}
1-\epsilon & 0 & 1-\epsilon & 0 \\
0 & \epsilon & 0 & \epsilon \\
\epsilon & 0 & \epsilon & 0 \\
0 & 1-\epsilon & 0 & 1-\epsilon
\end{pmatrix}
\end{align}

With the initial condition $p_0(y_0) \eqqcolon \begin{pmatrix}
  p_y^0\\
  1-p_y^0
\end{pmatrix} $, it follows:
\begin{align}
p_0(x_0,y_0) &= \begin{pmatrix}
1-\epsilon & 0 \\
0 & \epsilon  \\
\epsilon & 0  \\
0 & 1-\epsilon
\end{pmatrix}\, p_0(y_0)\\
&= \begin{pmatrix}
    (1-\epsilon)\,p_y^0\\
    \epsilon\, (1-p_y^0)\\
    \epsilon\,p_y^0\\
    (1-\epsilon)\, (1-p_y^0)
   \end{pmatrix}.
\end{align}

From this we can calculate all joint probabilities:
\begin{align}
 p(x_t,y_t) = \left( \mathbb{T}_x \mathbb{T}_y \right)^t \, p_0(x_0,y_0).
\end{align}

To evaluate the information flow [Eq.~\eqref{eqn_HMMDetachedEPx}], we need the joint probabilities $p(x_{t-1},y_t)$. These can be obtained from $p(x_{t-1},y_{t-1})$:
\begin{align}
 p(x_{t-1},y_t) = \mathbb{T}_y\,p(x_{t-1},y_{t-1}).
\end{align}

\begin{figure}[ht]
 \centering
\includegraphics[width = \linewidth]{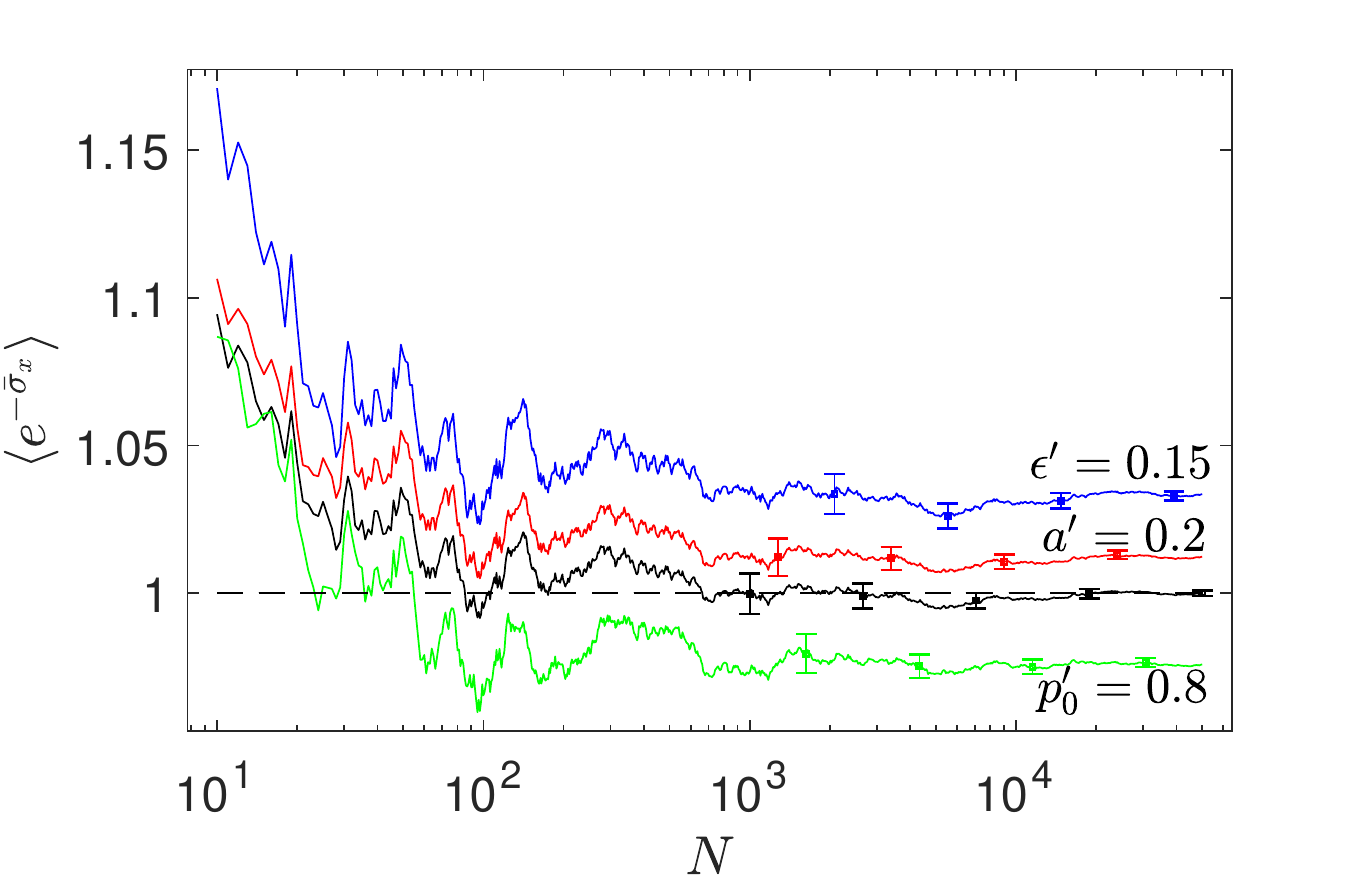}
 \caption{Convergence of the IFT for the auxiliary entropy production for our specific hidden Markov model. The trajectories were generated using the following parameters: $T = 5$, $a= 0.3$, $b= 0.2$, $p_0= 0.1$, and $\epsilon = 0.2$. We analyzed the IFT using the generating parameters (black line/ third from the top) and three sets of parameters in which one model parameter differs from the generating parameter (other lines). In the interest of clarity, error bars are only shown for larger sample sizes.}
 \label{fig_HMMConv}
\end{figure}

We demonstrate our formalism by generating $N$ trajectories of length $T$ with model parameters $a$, $b$, $p_0$ and $\epsilon$. We then calculate the auxiliary entropy production using Eq.~\eqref{eqn_defAuxEnt} for each trajectory and plot the left-hand-side of Eq.~\eqref{eqn_IFTAuxEnt} when evaluated with some other parameters $a'$, $b'$, $p_0'$ and $\epsilon'$. Figure.~\ref{fig_HMMConv} shows the result for a specific set of parameters.

One recognizes that only the correct set of parameters ensures convergence to one. This means that the IFT for the auxiliary entropy production can indeed be used to confirm the model parameters. 

In this specific hidden Markov model the IFT is more sensitive to the parameter $\epsilon$ governing the dynamics of the observed sequence than to the parameters $a$ and $b$.

\section{Discussion}\label{sec:discussion}
% What can be generalized from our results?
We have shown that within the field of stochastic thermodynamics measurement-feedback systems, sensors and hidden Markov models are related because they appear as special cases of a joint bipartite Markov chain.

This general joint bipartite Markov chain models a system with two interacting degrees of freedom. For such a case the influence of both subsystems as well as their interaction on the total entropy balance have to be considered. However, one can define other entropy productions like the detached entropy production which can be assigned a physical meaning.

From the study of the special cases we may deduce the following: whenever a subsystem is coupled to a thermal reservoir and the other subsystem influences its internal energy, its stochasticity is of thermal nature. Consequently, the detached entropy production equals the usual sum of system entropy change plus contribution of exchanged heat. On the other hand, when the influence is purely \emph{informational}, meaning it is due to a measurement, the detached entropy production takes the form of an information flow. This also holds true when the measurement noise is of thermal nature because it is not included in the energy budget of the joint system.

% What are methodological limitations?
We point out that the reverse process from which we derived the entropy production has a special structure because the subsystem updates swap order under time inversion. Generally, this fact is not important when the fluctuation theorems only involve forward quantities. It nonetheless plays a role if the parameter $\gamma$ in the detached IFT in Eq.~\eqref{eqn_detachedIFTx} differs from 0, since in that case $\gamma$ depends explicitly on the reverse detached probability.

% how do our results relate to other findings?
The dissection of the joint entropy production into detached entropy productions is similar to the partitioning of the joint entropy production rates into two sub-entropy productions with information flow~\cite{Horowitz2014}. Our formalism reveals the path probabilities (and reverse processes) from which one can define such sub-entropy productions.

Additionally, there is another way to decompose the joint entropy production which is close in spirit. Instead of detached entropy productions one may use \emph{marginal} entropy productions which are defined solely based on the marginal path probabilities. For example for $x$ one obtains:
\begin{align}
 \sigma_x \coloneqq \ln \frac{p_0(x_0)\,p_x(\bx|x_0)}{p_T(x_T)\,\bar p_x(\bbx|x_T)}.
\end{align}
The consequences of this separation have been studied by Crooks and Still~\cite{Crooks2016}. One finds that contrary to detached entropy production plus information flow, one obtains marginal entropy production plus transfer entropy. Both approaches seem equally valid. A comparison of the second-law inequalities these and other information measures provide has been made by Horowitz and Sandberg~\cite{Horowitz2014a}.

More than that, the decomposition of the joint process into two sub-processes is similar to the approach of causal conditioning within the information theory community. It is used to define directed information~\cite{Marko1973, Massey1990, Jiao2013} which is closely related to transfer entropy and has proven useful in the study of information thermodynamics~\cite{Ito2013, Hirono2015, Vinkler2016, Tanaka2017}.

The application of fluctuation theorems to validate the underlying model for data stemming from a stochastic process is a promising approach. It has already proven successful when estimating drift and diffusion coefficients in the Markov analysis of turbulent flows~\cite{Nickelsen2013, Reinke2016}.

% What are theoretical/practical implications?
Our findings point in two directions for future research. Firstly, it seems valuable to follow up on the study of how fluctuation theorems can be put to use to infer model parameters in, e.g., hidden Markov Models. It might be possible to use the fluctuation relation as a cost function in parameter learning or to infer the number of hidden parameters best describing the observed data.

Secondly, the question of how to appropriately describe systems which are strongly coupled to a thermal environment has recently gained attention~\cite{Seifert2016,Jarzynski2017,Miller2017}. A key issue is the partitioning into system and environment. Due to the strong coupling, perturbations of the environment due to the system may feed back into the future evolution of the system, thus violating the Markov assumption. We propose the detached path probabilities as a method to circumvent this problem and demonstrated that meaningful entropy measures can be derived from these probabilities.

\section{Conclusion}\label{sec:conclusion}
In this paper we used the detached path probabilities from which we defined the detached entropy production. This is a quantity which helps to analyze the stochastic thermodynamics of systems with interacting degrees of freedom.

We showed that one can understand measurement-feedback systems, sensors and hidden Markov models as special cases of one joint bipartite Markov chain. When applied to the special cases, the fluctuation relations involving the detached entropy production recover useful relations which have been found separately before.

Especially for hidden Markov models such a fluctuation relation can be used to confirm that model parameters have been learned correctly.
 
\begin{acknowledgments}
We thank Marcel Kahlen for a critical reading of the manuscript.
\end{acknowledgments}

\appendix*
\section{Posterior distribution for the hidden Markov model} \label{sec:A_posterior}

\subsection{Filtered marginals}
We begin by calculating the \textit{filtered marginals} $p(y_t|x_{0:t})$. These read:
\begin{align}
 p(y_t|x_{0:t}) &= \frac{p(x_{0:t},y_t)}{p(x_{0:t})} = \frac{p_x(x_t|y_t)\,p(y_t|x_{0:t-1})}{p(x_t|x_{0:t-1})},
\end{align}
where we have used the Markov property.

The second term in the numerator reads:
\begin{align}
 p(y_t|x_{0:t-1}) = \int d y_{t-1}\, p_y(y_t|y_{t-1})\, p(y_{t-1}|x_{0:t-1}).
\end{align}

This means that the current filtered marginal $p(y_t|x_{0:t})$ can be calculated recursively from the old filtered marginal $p(y_{t-1}|x_{0:t-1})$ when a new measurement $x_t$ comes in~\cite{Saerkkae2013}:
\begin{align}
 p(y_t|x_{0:t}) = \frac{p_x(x_t|y_t)\,\int d y_{t-1}\, p_y(y_t|y_{t-1})\, p(y_{t-1}|x_{0:t-1})}{\pi(x_t)},
\end{align}
where $\pi(x_t)$ ensures normalization. The recursion is started by:
\begin{align}
 p(y_0|x_0) = \frac{p_x(x_0|y_0)\, p_0(y_0)}{p_0(x_0)}.
\end{align}

\subsection{Posterior distribution}
At the conclusion of the process the procedure for the filtered marginal yields $p(y_T|x_{0:T})$. We are now interested in the posterior distribution $p(y_0,\by|x_0,\bx)$ of the hidden sequence given the entire observed sequence. We begin by noting:
\begin{align}
p(y_t|y_{t+1:T},x_{0:T}) &= p(y_t|y_{t+1},x_{0:t})\\
 &=  \frac{p_y(y_{t+1}|y_t)\,p(y_t|x_{0:t})}{p(y_{t+1}|x_{0:t})}.
\end{align}
Here, all probabilities can be obtained from the model and the filtered marginals. Now, we see that we can calculate the full posterior distribution recursively from the final filtered marginal $p(y_T|x_{0:T})$:
\begin{align}
 p(y_0,\by|x_0,\bx) = p(y_T|x_{0:T}) \, \prod\limits_{t=0}^{T-1} \frac{p_y(y_{t+1}|y_t)\,p(y_{t}|x_{0:t})}{p(y_t|x_{0:t-1})}.
\end{align}


\begin{thebibliography}{99}
\bibitem{FootnoteReverseDecomposition} Our definition of the reverse process does not preserve the order of subsystem updates. In the reverse process it is the $x$-subsystem which is first to change its state. For this reason the functions $\bar p(...;\bbu,\bbv)$ and $p(...;\bbu,\bbv)$ are not the same. However, for the detached probabilities  $\bar q_x(...)=q_x(...)$ holds so that the reversed detached probability is uniquely characterized by the reversed protocol. We keep the bar to be able to omit the dependency on the protocols later on.
\end{thebibliography}
\end{document}